\documentstyle[aps,epsf]{revtex}
\def\lya{{Ly $\alpha$}}
\def\gray{{$\gamma$-ray}}
\def\grays{{$\gamma$-rays}}
\def\ie{{\it i.e.}}
\def\eg{{\it e.g.}}
\def\etal{{\it et al.}}

\def\apj{{Astrophys. J.}}
\def\apjl{{Astrophys. J. (Lett.)}}
\def\apjs{{Astrophys. J. Supp.}}

\def\xxiicrc{{21st Internat. Cosmic Ray Conf.}}
\def\xxvicrc{{25th Internat. Cosmic Ray Conf.}}
\def\aap{{Astron. and Astr.}}
\def\mic{$\mu$m}
\def\aaps{{Astron. and Astr. Suppl.}}

\begin{document}

\baselineskip 14pt


\title{Extragalactic Absorption of High Energy Gamma-Rays}

\author{F.W. Stecker}

\address{Laboratory for High Energy Astrophysics, NASA Goddard Space Flight 
Center, Greenbelt, MD 20771, USA}


\maketitle

\begin{abstract}
The pair-production absorption of high-energy \grays\ by intergalactic low-energy photons is expected to produce a high-energy cutoff in the spectra of sources which is a sensitive function of redshift. We first discuss the
expected absorption coefficient as a function of energy and redshift derived by
Stecker and De Jager by making use 
of a new empirically based calculation of the spectral energy distribution of 
the intergalactic infrared radiation field as given by Malkan and Stecker.
We then discuss the fact that new data on the high energy \gray\ source Mrk 501
appear to show the amount of intergalactic absorption predicted. The 
implications of this new HEGRA data, should they be confirmed, are significant
for the astrophysics of this source, implying that (1) there is no significant
intrinsic absorption inside the source, and (2) the physics of the emission
spectrum produces a power-law to energies above 20 TeV. 
As a further test for intergalactic absorption, we give a predicted spectrum, 
with absorption included, 
for PKS 2155-304. This XBL lies at a redshift of 0.12, the highest 
redshift source yet observed at an energy above 0.3 TeV.
We also discuss the determination 
of the \gray\ opacity at higher redshifts (out to $z=3$), following the 
treatment of Salamon and Stecker.

\end{abstract}


\section{Introduction}

Very high energy \gray\ beams from blazars can be used to 
measure the intergalactic infrared radiation field, since 
pair-production interactions of \grays\ with intergalactic IR photons 
will attenuate the high-energy ends of blazar spectra \cite{sds92}. 
In recent years, this concept has been used successfully to place upper limits 
on the the intergalactic IR field (IIRF) \cite{sd93}
- \cite{bil98}.
Determining the (IIRF), in turn, allows us to 
model the evolution of the galaxies which produce it. 
As energy thresholds are lowered 
in both existing and planned ground-based
air Cherenkov light detectors \cite{knp}, cutoffs in the \gray\ spectra of 
more distant blazars are expected, owing to extinction by the IIRF. These
can be used to explore the redshift dependence of the 
IIRF \cite{ss97}, \cite{ss98}. 

There are now 66 ``grazars'' ($\gamma$-ray blazars) which have been 
detected by the {\it EGRET} team
\cite{3egret}. These sources, optically violent variable quasars
and BL Lac objects, have been detected out to a redshift greater that 2.
Of all of the blazars detected by {\it EGRET}, only the low-redshift 
BL Lac, Mrk 421 ($z = 0.031$), has been seen by
the Whipple telescope \cite{punch92}. The fact that the Whipple team did not 
detect the much brighter {\it EGRET} source, 3C279, at TeV energies
\cite{vac90}, \cite{ker93} is consistent with the predictions of a
cutoff for a source at its much higher redshift of 0.54 \cite{sds92}.
So too are the further detections of three other close BL Lacs 
($z < 0.12$), {\it viz.}, Mrk 501 ($z = 0.034$) \cite{quinn96}, 1ES2344+514 
($z = 0.044$)\cite{cat98}, and PKS 2155-304 ($z = 0.117$) \cite{chad98} which 
were too faint at GeV energies to be seen by {\it EGRET}\footnotemark\footnotetext{PKS 2155-304 was seen in one observing period by
{\it EGRET} as reported in the Third EGRET Catalogue \cite{3egret}}.


The formulae relevant to absorption calculations involving pair-production 
are given and discussed in Ref. \cite{sds92}.
For $\gamma$-rays in the TeV energy range, the pair-production cross section 
is maximized when the soft photon energy is in the infrared range:
\begin{equation} 
\lambda (E_{\gamma}) \simeq \lambda_{e}{E_{\gamma}\over{2m_{e}c^{2}}} =
2.4E_{\gamma,TeV} \; \; \mu m 
\end{equation}
where $\lambda_{e} = h/(m_{e}c)$ 
is the Compton wavelength of the electron.
For a 1 TeV $\gamma$-ray, this corresponds to a soft photon having a
wavelength  near the K-band (2.2\mic). (Pair-production interactions actually
take place with photons over a range of wavelengths around the optimal value as
determined by the energy dependence of the cross section; see eq. (3).) 
If the emission spectrum of
an extragalactic source extends beyond 20 TeV, then the extragalactic
infrared field should cut off the {\it observed} spectrum between $\sim
20$ GeV and $\sim 20$ TeV, depending on the redshift of the source \cite{ss97},
\cite{ss98}.

\section{Absorption of Gamma-Rays at Low Redshifts}

Stecker and De Jager \cite{sd98} (hereafter SD98) have recalculated the 
absorption coefficient of intergalactic
space using a new, empirically based calculation
of the spectral energy distribution (SED) of intergalactic low energy 
photons by Malkan and Stecker \cite{ms98} (hereafter MS98) 
obtained by integrating luminosity dependent infrared spectra of galaxies
over their luminosity and redshift distributions.
After giving their results on the \gray\ optical depth as a function of energy 
and redshift out to a redshift of 0.3, SD98 applied their calculations by
comparing their results with the spectral data on Mrk 421 \cite{mch97} and 
spectral data on Mrk 501 \cite{ah98}. 

SD98 make the reasonable simplifying assumption 
that the IIRF is basically in
place at a redshifts $<$ 0.3, having been produced primarily at higher
redshifts \cite{mad99}. 
Therefore SD98 limited their calculations to $z<0.3$. (The calculation of
\gray\ opacity at higher redshifts \cite{ss97},\cite{ss98} will be discussed
in the next section.)

\begin{figure}
\vspace{1.0truecm}
\epsfysize=4.0 truein
\centerline{\epsfbox{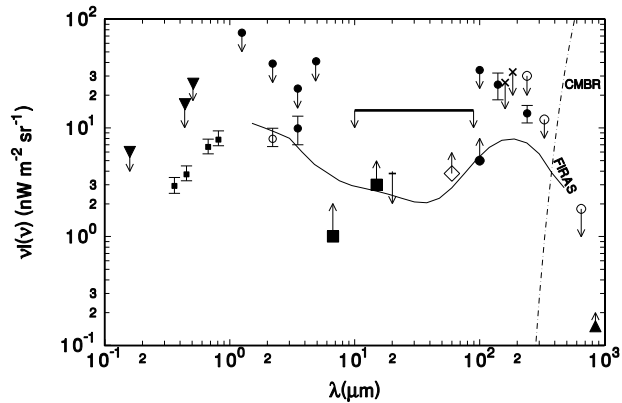}}
\vspace{-5.0truecm}
\caption{{The upper infrared SED from Malkan and Stecker compared with 
observational data and other constraints (courtesy O.C. De Jager).}}
\label{irdata}
\end{figure}

SD98 assumed for the IIRF, two of the SEDs given in MS98 \cite{ms98}. Their
upper curve now appears to be in better agreement with
lower limits from galaxy counts, with Keck telescope, {\it HST}. {\it NICMOS},
{\it ISO} and {\it SCUBA} studies of galaxies
at high redshifts (Ref.\cite{mad99} and references therein) and with 
{\it COBE} data \cite{pug96}  - \cite{lag99} (see Figure \ref{irdata}).

~

The results of MS98 are also in agreement with upper limits obtained from
TeV \gray\ studies \cite{sd93} - \cite{bil98}.
This agreement is illustrated in Figure \ref{irdata} which shows the 
upper SED curve from MS98 in comparison with various data and limits.

The SD98 results for the absorption coefficient as 
a function of energy do not differ dramatically from those obtained
previously \cite{mp96}, \cite{sd97}; 
however, they are more reliable because they are
based on the empirically derived IIRF given by MS98, whereas all previous
calculations of TeV $\gamma$-ray absorption were based on theoretical modeling
of the IIRF. 
The MS98 calculation was based on data from nearly 3000
IRAS galaxies. These data included (1) the luminosity dependent infrared SEDs
of galaxies, (2) the 60$\mu$m luminosity function of galaxies and, (3) 
the redshift distribution of galaxies.

\begin{figure}
\vspace{2.0truecm}
\epsfysize=4.0 truein
\centerline{\epsfbox{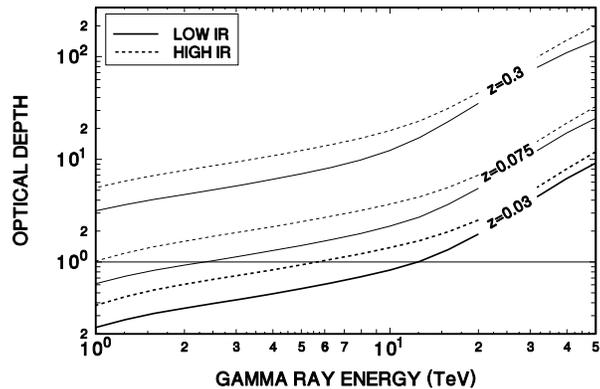}}
\vspace{-5.0truecm}
\caption{{Optical depth versus energy for \grays\ originating at various 
redshifts obtained using the SEDs corresponding to the lower IIRF
(solid lines) and higher IIRF (dashed lines) levels shown in a Figure 
taken from SD98. As discussed in the text, the higher IIRF curves 
(dashed lines) are more in line with recent data.}}
\label{lowztau}
\end{figure}

The advantage of using empirical
data to construct the SED of the IIRF, as done in MS98, is 
particularly indicated in the mid-ir range where
galaxy observations indicate more flux from warm dust in galaxies than that
taken account of in more theoretically oriented models. As a consequence, 
the mid-IR ``valley'' between the cold dust 
peak in the far IR and cool star peak in the near IR is partially filled in 
(see Figure \ref{irdata}).
For a source at low redshift, it follows from eq. (1) that \grays\ of 
energy $\sim$ 20 TeV will be absorbed preferentially by photons in the
wavelength range of this ``valley'', \ie, near 50 \mic. In this 
range, significant lower limits now exist which are near the predicted
IIRF flux (see Figure \ref{irdata}).

\begin{figure}
\epsfysize=4.0 truein
\centerline{\epsfbox{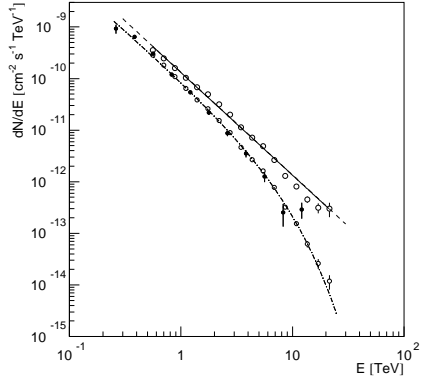}}
\vspace{-2.0truecm}
\caption{{The bottom curve and points show the new HEGRA data on Mrk 501 
in the flaring state \protect\cite{kono1}; the upper line and points show 
the intrinsic spectrum of the source with the effect of the predicted 
extragalactic absorption removed \protect\cite{kono2}. }}
\label{hegra}
\end{figure}

In fact, the observed flaring spectrum of Mrk 501 has been 
newly extended to an energy of 24 TeV by observations of the {\it HEGRA} group
\cite{kono1}. 
The new {\it HEGRA} data are well fitted
by an $E^{-2}$ source spectrum steepened at
energies above a few TeV by intergalactic absorption with the optical depth
calculated by SD98 \cite{kono2}. Figure \ref{hegra}, taken from
Ref. \cite{kono2}, clearly shows this. The philosophy behind 
Ref. \cite{kono2} is
that the existing lower limits on the mid-ir background flux predict a minimum
expected absorption. The derived unabsorbed source spectrum then tells us
(1) that there is negligible intrinsic absorption in the source, and (2) the
physics of the emission mechanism should give a power-law spectrum with a
spectral index of $\sim$2 up to an energy of at least $\sim$ 20 TeV.

\begin{figure}
\epsfysize=3.0 truein
\hspace{3truecm}
\epsfbox{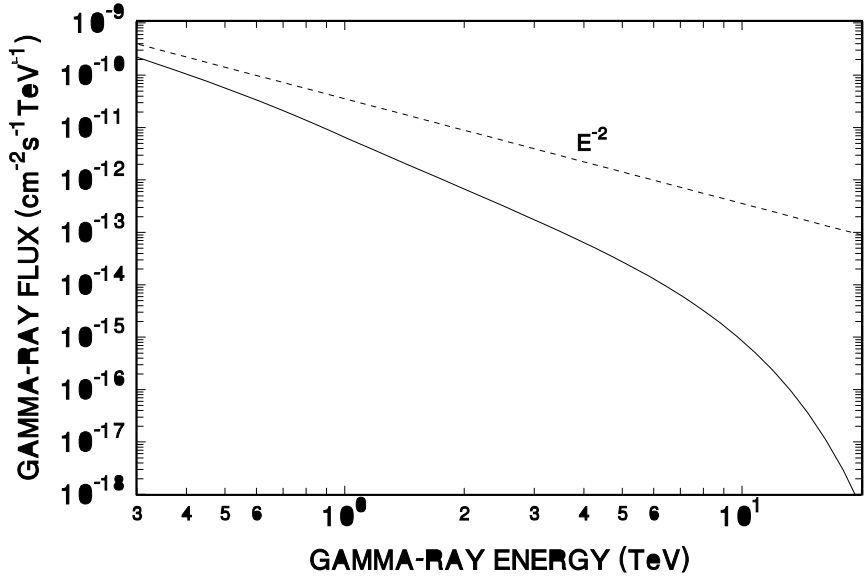}
\vspace{1truecm}
\caption{{Predicted differential absorbed spectrum, for PKS 2155-304 
(solid line) assuming an $E^{-2}$ differential source spectrum (dashed line) 
normalized to the integral flux given in Ref. \protect\cite{chad98}
(see text).}}
\label{2155}
\end{figure}

Consider the source PKS 2155-304, an XBL located at a moderate
redshift of 0.117, which has been reported by the Durham group to have
a flux above 0.3 TeV of $\sim 4 \times 10^{-11}$ cm$^{-2}$ s$^{-1}$
\cite{chad98}. We predict that this source should 
have its spectrum steepened by $\sim$ 1 in its spectral
index between $\sim 0.3$ and $\sim 3$ TeV and should show an absorption 
turnover above $\sim 6$ TeV as shown in Figure
\ref{2155}. Observations of the spectrum of this source should
provide a further test for intergalactic absorption.

\section{Absorption of Gamma-Rays at High Redshifts}
 
In order to calculate high-redshift absorption properly, it is
necessary to determine the spectral distribution of the intergalactic low 
energy photon background radiation as a function of redshift as realistically 
as possible out to frequncies beyond the Lyman limit. This calculation,
in turn, requires observationally based information on the evolution of the 
spectral energy distributions (SEDs) of IR through UV starlight from galaxies,
particularly at high redshifts. 

Salamon and Stecker \cite{ss98} (hereafter SS98) have calculated 
the \gray\ opacity as a function of both energy and redshift
for redshifts as high as 3 by taking account of the evolution of both the
stellar population spectra and emissivity of galaxies with redshift. 
In order to accomplish this, 
they adopted the recent analysis of Fall, \etal\  \cite{fall96} and 
also included the effects of metallicity evolution on galactic stellar 
population spectra.
They then gave predicted \gray\ spectra 
for selected blazars and extend our calculations of the 
extragalactic \gray\ background from blazars to an energy of 500 GeV with
absorption effects included.

Fall, \etal \cite{fall96} have devised a method for calculating stellar 
emissivity which bypasses the uncertainties associated with estimates of 
poorly defined luminosity distributions of evolving galaxies.
The core idea of their 
approach is to relate the star formation rate 
directly to the evolution of the neutral gas density in damped
\lya\ systems, and then to use stellar population synthesis models to
estimate the mean co-moving stellar emissivity ${\cal E}_{\nu}(z)$
of the universe as a function of frequency $\nu$ and
redshift $z$.

The SS98 calculation of stellar emissivity closely follows this 
elegant analysis, with minor modifications.
SS98 also obtained metallicity correction
factors for stellar radiation at various wavelengths. Decreased metallicity at
high redshifts gives a bluer stellar population spectrum \cite{wo94}, 
\cite{be94}.

The stellar emissivity in the universe is found to peak
at $ 1 \le z \le 2$, dropping off steeply at lower redshifts and more slowly
at higher redshifts. Indeed, observational data from the 
Hubble Deep Field to show that metal production has a similar redshift 
distribution, such production being a direct measure of the star formation 
rate (see, \eg, Ref.\cite{mad99}).

With the co-moving energy density $u_{\nu}(z)$ evaluated \cite{ss98} (SS98), 
the optical depth for \grays\ owing to electron-positron pair production 
interactions with photons of the stellar radiation
background can be determined from the expression \cite{sds92}

\begin{equation} \label{G}
\tau(E_{0},z_{e})=c\int_{0}^{z_{e}}dz\,\frac{dt}{dz}\int_{0}^{2}
dx\,\frac{x}{2}\int_{0}^{\infty}d\nu\,(1+z)^{3}\left[\frac{u_{\nu}(z)}
{h\nu}\right]\sigma_{\gamma\gamma}(s)
\end{equation}
where $s=2E_{0}h\nu x(1+z)$,
$E_{0}$ is the observed \gray\ energy at redshift zero, 
$\nu$ is the frequency at redshift $z$,
$z_{e}$ is the redshift of
the \gray\ source, $x=(1-\cos\theta)$, 
and the pair production cross section $\sigma_{\gamma\gamma}$ is zero for
center-of-mass energy $\sqrt{s} < 2m_{e}c^{2}$, $m_{e}$ being the electron
mass.  Above this threshold, 
\begin{equation} \label{H}
\sigma_{\gamma\gamma}(s)=\frac{3}{16}\sigma_{\rm T}(1-\beta^{2})
\left[ 2\beta(\beta^{2}-2)+(3-\beta^{4})\ln\left(\frac{1+\beta}{1-\beta}
\right)\right],
\end{equation}
where $\beta=(1-4m_{e}^{2}c^{4}/s)^{1/2}$.

\begin{figure}
\vspace{1.0truecm}
\epsfysize=3.0 truein
\centerline{\epsfbox{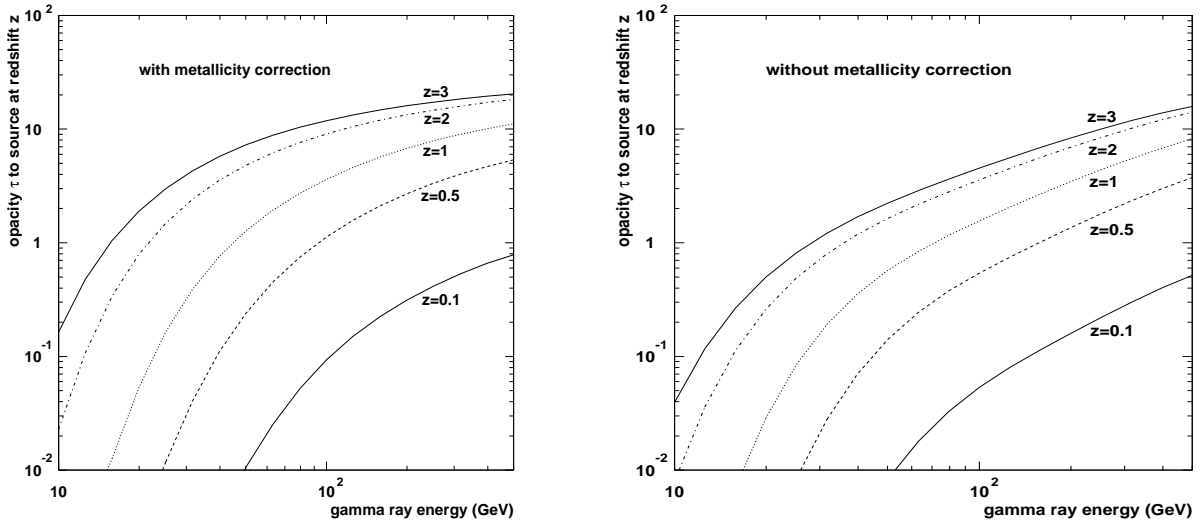}}
\caption{{ The opacity $\tau$ of the universal soft photon background to
\grays\ as a function of \gray\ energy and source redshift (from SS98)
\protect\cite{ss98}. These curves are calculated with and without a
metallicity correction.
}}
\label{opac}
\end{figure}

Figure \ref{opac}
shows the opacity $\tau(E_{0},z)$ for the energy
range 10 to 500 GeV, calculated by SD98 both with and without a 
metallicity correction.
Extinction of \grays\ is negligible below 10 GeV.
The weak redshift dependence of the opacity at the higher redshifts 
as shown in Figure ~\ref{opac}  indicates that the opacity is not very
sensitive to the initial epoch of galaxy formation, $z_{max}$. In fact, the 
uncertainty in the
metallicity correction (see Figure \ref{opac}) would obscure any dependence on
$z_{max}$ even further.

\section{The Effect of Absorption on the Spectra of Blazars and Gamma-ray
Bursts}
 
\begin{figure}
\vspace{1.0truecm}
\epsfysize=3.5 truein
\centerline{\epsfbox{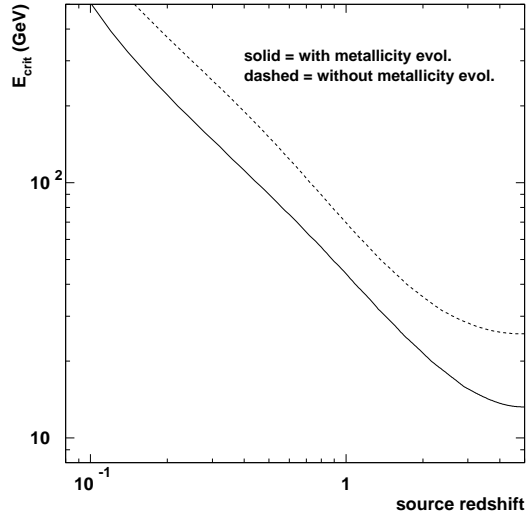}}
\caption{{ The critcal energy for \gray\ absorption above which the optical
depth is predicted to be greater than 1 as a function of the redshift of
the source (from the results of SS98)
\protect\cite{ss98} (see text).}}
\label{ecritvsz}
\end{figure}

With the \gray\ opacity $\tau(E_{0},z)$ calculated out to
$z=3$,
the cutoffs in blazar \gray\ spectra caused by extragalactic pair 
production interactions with stellar photons can be predicted.
Figure \ref{ecritvsz}, based on the results given in 
Ref. \cite{ss98} (SS98),
shows the expected effect of the intergalactic radiation grazar 
and \gray\ burst spectra. 
This figure plots the critical energy for absorption
(\ie, for $\tau = 1$) versus redshift. For energies much above the critical 
energy, the optical depth is greater than 1, leading to a predicted cutoff in
the spectrum of the extragalactic source.

The discovery of optical and X-ray afterglows of \gray\ bursts and 
the identification of host galaxies with measured redshifts (see, \eg,
Refs. \cite{me97} and \cite{ku98}) has lead the accumulation of evidence that
these bursts are highly relativistic fireballs originating at cosmological
distances \cite{liv98} and may be associated primarily with early star
formation \cite{dj98}.

As indicated in Figure \ref{ecritvsz}
\grays\ above an energy of $\sim$ 15 GeV will
be attenuated if they at emitted at a redshift of $\sim$ 3.
On 17 February 1994, the {\it EGRET} telescope observed a \gray\ burst
which contained a photon of energy $\sim$ 20 GeV \cite{hu94}.
As an example, if one adopts the opacity results which include the
metallicity correction, the highest energy photon in this burst 
would be constrained probably
to have originated at a redshift
less than $\sim$2. 
Future detectors such as {\it GLAST} Ref. \cite{bl96}, may be able to 
place better redshift constraints on bursts observed at higher energies. 
Such constraints may further help to identify the host galaxies of \gray\
bursts.

Observed cutoffs in grazar spectra 
may be intrinsic cutoffs in \gray\ production in
the source, or may be caused
by intrinsic \gray\ absorption within the source itself. In fact, models of
quasar emission can predict natural cutoffs in quasar emission spectra in the 
relevant energy range above $\sim$ 10 GeV. 
Whether or not cutoffs in grazar spectra are 
primarily caused by intergalactic absorption can be determined by
observing whether the grazar cutoff energies have the 
type of redshift dependence predicted here.

\section{Acknowledgment}

The work presented here was a result of extensive 
collaboration with O.C. De Jager, M.A. Malkan, and M.H. Salamon, as indicated 
in the references cited.

\end{document}